\documentclass[conference]{IEEEtran}
\IEEEoverridecommandlockouts

\usepackage{cite}
\usepackage{amsmath,amssymb,amsfonts}
\usepackage{algorithmic}
\usepackage{graphicx}
\graphicspath{{Figures/}} 
\usepackage{subcaption}
\usepackage{multirow}
\usepackage{caption}
\usepackage[font=footnotesize]{caption}

\usepackage{cite}
\usepackage{csquotes}

\usepackage{textcomp}
\usepackage{xcolor}
\def\BibTeX{{\rm B\kern-.05em{\sc i\kern-.025em b}\kern-.08em
 T\kern-.1667em\lower.7ex\hbox{E}\kern-.125emX}}

\usepackage{stfloats}
\usepackage{hyperref}
\hypersetup{
	colorlinks=true,
	linkcolor=blue,
	citecolor=blue,
	urlcolor=blue
}

\makeatletter
\newcommand{\linebreakand}{%
  \end{@IEEEauthorhalign}
  \hfill\mbox{}\par
  \mbox{}\hfill\begin{@IEEEauthorhalign}
}
\makeatother

\begin{document}
\title{
Decentralized Bus Voltage Restoration\\ for DC Microgrids
}

\author{
\IEEEauthorblockN{Nabil Mohammed, Shehab Ahmed, Charalambos Konstantinou}
\IEEEauthorblockA{Computer, Electrical and Mathematical Science and Engineering (CEMSE) Division\\
Center for Renewable Energy and Storage Technologies (CREST)\\
King Abdullah University of Science and Technology (KAUST)\\
Thuwal, Saudi Arabia\\
E-mail: \{nabil.mohammed, shehab.ahmed, charalambos.konstantinou\}@kaust.edu.sa}
}

\maketitle

\begin{abstract}
Regulating the voltage of the common DC bus,  also referred to as the \enquote{load bus},  in DC microgrids is crucial for ensuring reliability and maintaining the nominal load voltage, which is essential for protecting sensitive loads from voltage variations. Stability and reliability are thereby enhanced, preventing malfunctions and extending the lifespan of sensitive loads (e.g., electronic devices). Voltage drops are caused by resistances of feeders connecting converters to the common DC bus, resulting in a reduced DC bus voltage compared to the nominal/desired value. Existing techniques to restore this voltage in DC microgrids are mainly centralized and rely on secondary control layers. These layers sense the common DC bus voltage, compare it to the nominal value, and utilize a PI controller to send corrections via communication links to each converter. In this paper, a local and straightforward approach to restoring the bus voltage in DC microgrids is presented, ensuring regulation in a decentralized manner. Voltage drops across resistances of feeders  are compensated by an additional control loop feedback within each converter, based on the converter output current and feeder resistance. The proposed approach is verified through simulation and hardware-in-the-loop results, eliminating the need for communication links and hence increasing reliability and reducing cybersecurity threats.
\end{abstract}


\section{Introduction}\label{Sec1}
In modern converter-dominated power systems, the integration of emerging distributed generation technologies has become increasingly prevalent, with solar photovoltaic (PV) and wind energy widely adopted. Microgrids, which aggregate various sources and loads into a single dispatchable system, can operate either interconnected with the main grid (AC or DC) or in standalone mode \cite{hatziargyriou2007microgrids, pogaku2007modeling}. They are mainly categorized into three types: AC microgrids, DC microgrids, and hybrid microgrids that combine both AC and DC elements. Unlike the complex control requirements of AC microgrids, which include managing frequency, voltage, and reactive power, DC microgrids are preferable in several applications due to their simpler control mechanisms \cite{guerrero2010hierarchical, mohammed2023accurateVOC, 10542752}.

DC microgrids can be further classified based on control configurations into centralized, distributed, and decentralized architectures. Centralized control is characterized by a single control center responsible for decision-making, which simplifies management but introduces a single point of failure risk and scalability challenges. Distributed control employs a hierarchical structure that combines centralized and decentralized elements, balancing optimization and resilience but requiring complex implementation and robust communication infrastructure. In decentralized control, independent controllers make local decisions, enhancing reliability and scalability, though coordination and integration are more complex. The selection of control architecture is determined by the specific needs and infrastructure of the microgrid \cite{dragivcevic2015dc}.

In DC microgrids, the primary control objectives are load power sharing and DC bus voltage regulation. The resistances of the feeders connecting the converters to the common DC bus play a crucial role in achieving these objectives. On the one hand, the load power sharing is typically ensured via a droop control strategy, which is implemented at the converter level and is effective under conditions of identical or matched feeder resistances. However, when feeder resistances are nonidentical or mismatched, additional control solutions must be adopted. These solutions include adaptive droop coefficients and virtual resistances, which are necessary to maintain accurate load sharing \cite{nasirian2014distributed,guerrero2010hierarchical, mohammed2023accurate}.

On the other hand, voltage deviation in DC microgrids refers to fluctuations from the desired voltage levels, primarily caused by line resistances due to inherent conductor resistance and high current flows, leading to voltage drops and power losses. Several voltage restoration techniques are employed in the literature \cite{lu2013improved, dragivcevic2015dc, guerrero2010hierarchical, montegiglio2023decentralized, wang2024adaptive}. Existing techniques to restore voltage in DC microgrids are commonly centralized and rely on a secondary control layer. This layer senses the DC bus voltage, compares it to the nominal value, and utilizes a PI controller to send corrections via communication links to each converter. However, reliance on communication links makes both centralized and distributed control approaches to be potentially vulnerable to cyber attacks and failures when communication links fail \cite{zhou2020cyber}.

This paper presents a decentralized method for restoring bus voltage in DC microgrids. The proposed method compensates for voltage drops across each feeder line by implementing an additional control loop feedback within each converter, utilizing only the converter output current and feeder resistance. Unlike centralized methods that require secondary control loops and communication links to restore voltage to the nominal value, the proposed method is local, thereby obviating the necessity for communication links, enhancing reliability, and mitigating cybersecurity risks.

The remainder of this paper is organized as follows: Section~\ref{Sec_2} provides an overview of traditional centralized islanded DC Microgrids. Section~\ref{Sec_3} presents the proposed decentralized methodology for restoring bus voltage in DC microgrids. Section~\ref{Sec_4} reports simulation and hardware-in-the-loop verification results. Finally, Section~\ref{Sec_5} concludes the paper.

\section{Traditional Centralized Islanded\\ DC Microgrids}\label{Sec_2}
\begin{figure*}[!t]
  \centering
  \includegraphics [width=0.85\linewidth]{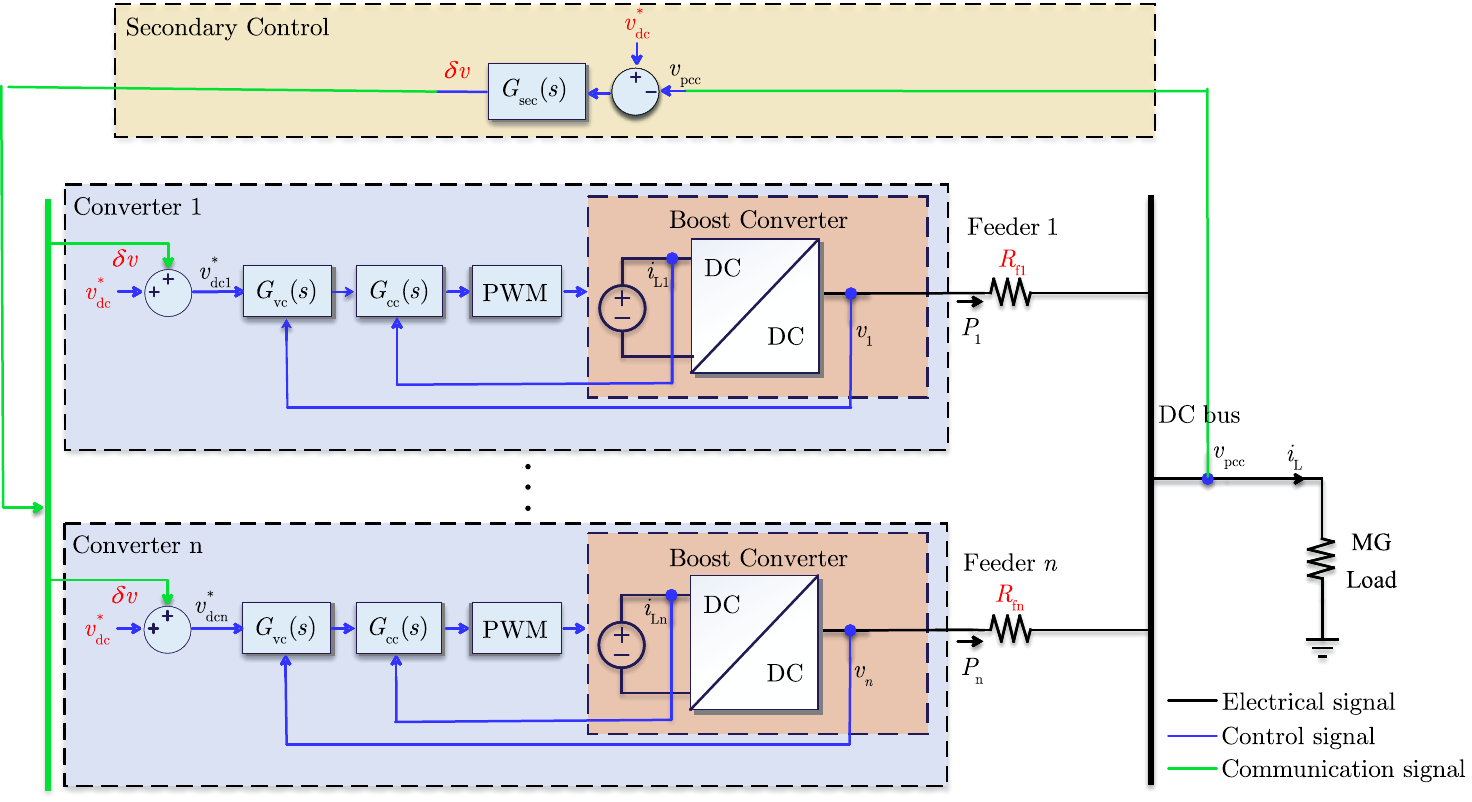}
   \caption{Traditional centralized control (with secondary control loop) for DC bus voltage restoration in DC
microgrids.} 
  \label{Fig_Meth_Trad_cont}
\vspace{-0.350cm}
\end{figure*}
The general structure of a centralized DC microgrid is illustrated in Fig.~\ref{Fig_Meth_Trad_cont}. The microgrid consists of $n$ parallel boost converters, MG load, and a secondary control loop. Each boost converter is equipped with a local current controller ($G_\mathrm{cc}(s)$) and voltage controller ($G_\mathrm{vc}(s)$). Converter 1 regulates the output voltage to a reference value ($v^*_\mathrm{dc1}$, and similarly to $v^*_\mathrm{dcn}$ for converter $n$). 
The resistances $R_\mathrm{f1}$ to $R_\mathrm{fn}$ of the feeders connect converter 1 to converter $n$ to the common DC bus, respectively. In this structure, the MG load is aggregated at the DC bus, resulting in a total load current of $i_\mathrm{L}$.

When the converters and the feeder resistances are identical, the current supplied by each converter to the load can be expressed as:

\begin{equation}
\begin{aligned}
i_1 &= \frac{v_1 - i_\mathrm{L} R_\mathrm{L}}{R_\mathrm{f1}}, \\
i_n &= \frac{v_n - i_\mathrm{L} R_\mathrm{L}}{R_\mathrm{fn}}, \\
i_1 &= i_n = \frac{i_\mathrm{L}}{n}.
\label{eq1}
\end{aligned}
\end{equation}
From \eqref{eq1}, it is evident that the current supplied to the load by each converter (e.g., $i_1$) is directly impacted by the resistance value of its feeder ($R_\mathrm{f1}$). The feeder resistances also cause voltage drops across the feeders, calculated as follows:

\begin{equation}
\begin{aligned}
\Delta v_\mathrm{f1} &= R_\mathrm{f1} i_1, \\
\Delta v_\mathrm{fn} &= R_\mathrm{fn} i_n.
\label{eq2}
\end{aligned}
\end{equation}

As shown in \eqref{eq2}, the voltage of the DC bus at the point-of-common-coupling (PCC) ($v_\mathrm{pcc}$) will always be below the desired nominal value ($v^*_\mathrm{dc}$). To restore the voltage to the nominal value, the secondary control loop is employed. This involves continuous sensing of the DC bus voltage, comparing it to the nominal value, and using a PI controller to send corrections via communication links to each converter. As depicted in Fig.~\ref{Fig_Meth_Trad_cont}, the control law for voltage restoration of converter 1 and converter $n$, similar to the other converters, considering the secondary control loop is expressed as:

\begin{equation}
\begin{aligned}
v^*_\mathrm{dc1} = v^*_\mathrm{dc} + (K_\mathrm{p-sec} + \frac{K_\mathrm{i-sec}}{s})(v^*_\mathrm{dc} - v_\mathrm{pcc}), \\
v^*_\mathrm{dcn} = v^*_\mathrm{dc} + (K_\mathrm{p-sec} + \frac{K_\mathrm{i-sec}}{s})(v^*_\mathrm{dc} - v_\mathrm{pcc}),\\
\label{eq3}
\end{aligned}
\end{equation}
where $K_\mathrm{p-sec}$ and $K_\mathrm{i-sec}$ are the proportional and integral coefficients of the secondary PI compensator to restore the DC bus voltage measured at the PCC to the desired nominal value, $v^*_\mathrm{dc}$.

The primary challenge in \eqref{eq3} is the complete dependence of the secondary (centralized) control on communication links. This dependency impacts MG reliability due potential issues or/and vulnerabilities related to communication failures and cyber attacks in practical applications.

\section{Proposed Decentralized Islanded DC Microgrids}\label{Sec_3}
\begin{figure*}[!t]
  \centering
  \includegraphics[width=0.75\linewidth]{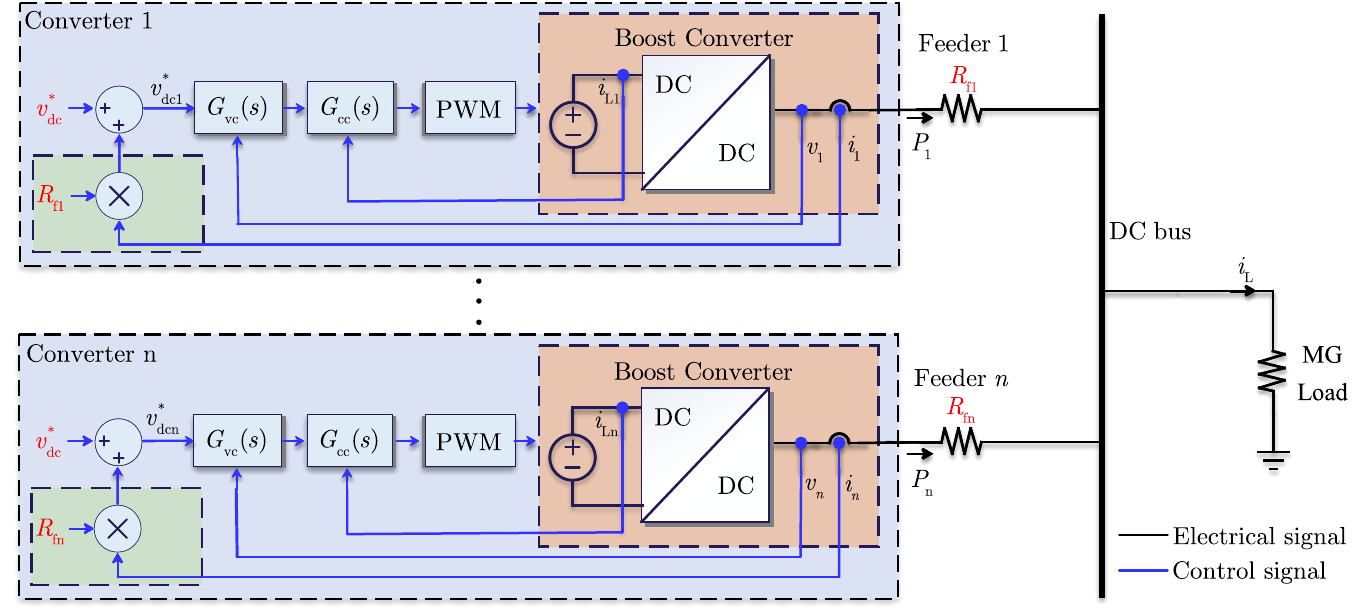}
  \caption{Proposed decentralized control (without secondary control loop) for DC bus voltage restoration in DC microgrids.} 
  \label{Fig_Meth_Prop_cont}
  \vspace{-0.350cm}
\end{figure*}
Fig.~\ref{Fig_Meth_Prop_cont} illustrates the control architecture for the DC microgrid employing the proposed decentralized control strategy. In this approach, the secondary control loop shown in Fig.~\ref{Fig_Meth_Trad_cont}, along with all associated communication links, is eliminated.

The proposed strategy utilizes a local control approach to restore bus voltage in DC microgrids, ensuring regulation in a decentralized manner. Voltage drops across feeder lines are compensated by additional control loop feedback within each converter, based on the converter output current and its feeder resistance. This decentralized control method ensures effective and automated DC bus voltage regulation, making it suitable for fully decentralized DC microgrids.

Drawing upon Fig.~\ref{Fig_Meth_Trad_cont} and \eqref{eq2}, the control law for the proposed decentralized control of converter 1 and converter $n$, similar to the other converters, for DC bus voltage restoration in DC microgrid is expressed as:

\begin{equation}
\begin{aligned}
v^*_\mathrm{dc1} = v^*_\mathrm{dc} + i_\mathrm{1}R_\mathrm{f1}, \\
v^*_\mathrm{dcn} = v^*_\mathrm{dc} + i_\mathrm{n}R_\mathrm{fn}.
\label{eq4}
\end{aligned}
\end{equation}
In \eqref{eq4}, the values of the $n$ resistances can either be entered using known parameters of the feeders or estimated online \cite{mohammed2023accurate}.
By comparing the traditional centralized control presented in \eqref{eq3} with the proposed decentralized control strategy presented in \eqref{eq4}, it is noted that voltage restoration can be achieved accurately without requiring a secondary controller, thereby enhancing the reliability of the microgrid system.

\begin{table}[t!]
	\renewcommand{\arraystretch}{1.3}
	\caption{Parameters used for the simulation study.}
	\centering
	\label{table_1}
	    \resizebox{8cm}{!}{
		\begin{tabular}{l l l l}
			\hline\hline \\[-3mm]
			{Quantity} & {Symbol}& {Value}  &{Unit}  \\[1.6ex] \hline
{Microgrid}   &&&\\
{Nominal voltage}&$v^*_\mathrm{dc}$ & 400&V  \\
{Full load current}&$i_\mathrm{L} $ & 30  &kA \\
{Resistance of feeders 1,2 and 3}&{$R_\mathrm{f1}$, $R_\mathrm{f2}$, $R_\mathrm{f3}$ }& {0.4} &{$\Omega$} \\
\hline
{DC boost converters 1, 2 and 3}   &&&\\
{Rated voltage} &$ v_\mathrm{rated} $& 400 &V  \\
{Rated power}&$P_\mathrm{rated} $& 5 &kW \\
{Rated current}&$i_\mathrm{rated} $& 12.5 &A \\
Switching frequency & $f_\mathrm{sw}$& 5 &kHz \\
\hline
\multicolumn{4}{l}{Current and voltage controllers for individual converter}
\\
{Inter current controller}   &&&\\
{Proportional gain}&$ K_\mathrm{p-cc}$ & {1}& \\
{Integral gain}&$ K_\mathrm{i-cc}$ & {500}&1/sec \\
{Inter voltage controller}   &&&\\
{Proportional gain}&$ K_\mathrm{p-vc}$ & {7}& \\
{Integral gain}&$ K_\mathrm{i-vc}$ & {100}&1/sec \\
\hline
\multicolumn{4}{l}{Secondary control of the {microgrid}}\\
{Proportional gain}&$ K_\mathrm{p-sec}$ & {1.5}& \\
{Integral gain}&$ K_\mathrm{i-sec}$ & {150}&1/sec \\ [1.4ex]
			\hline\hline
		\end{tabular}
	}
\vspace{-0.3cm}
\end{table}

\begin{figure}[!t]
    \centering
    \subfloat[\label{fig3_a}]{\includegraphics [width=0.990\linewidth]{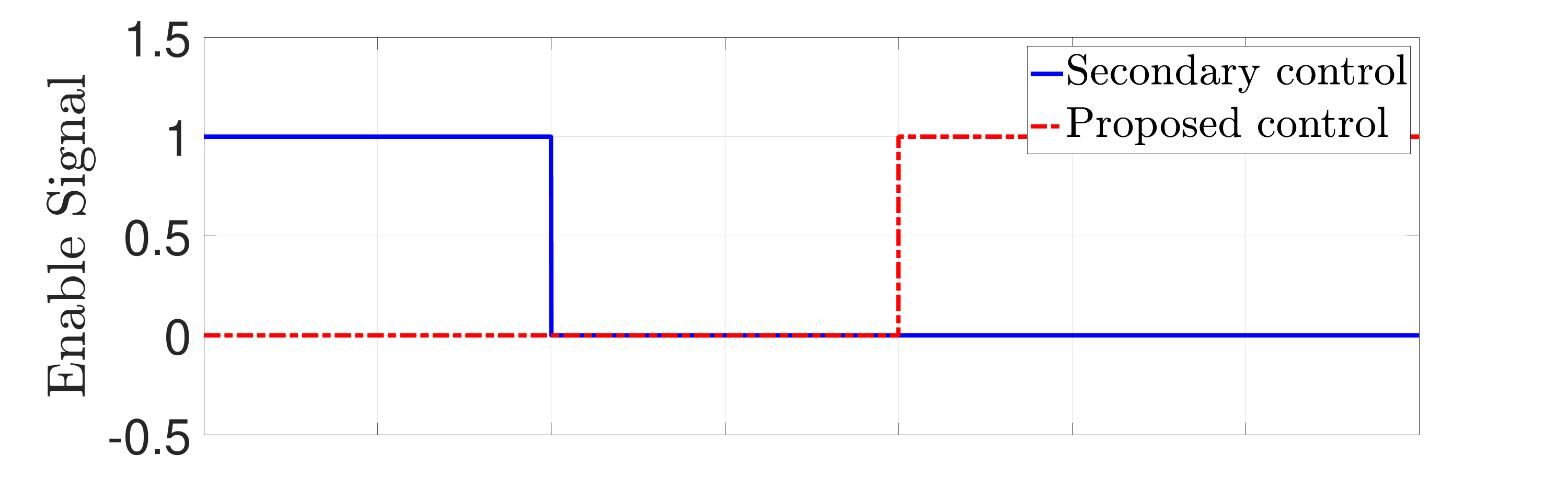}}\\ 
    \subfloat[\label{fig3_b}]{\includegraphics [width=0.990\linewidth]{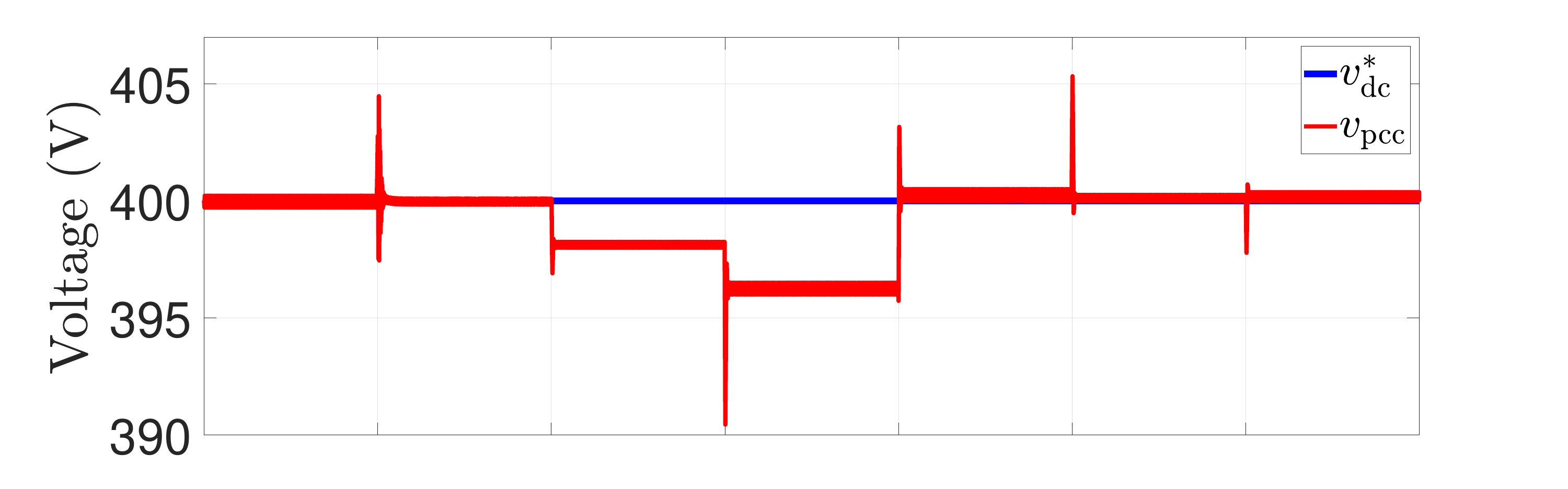}}\\
    \subfloat[\label{fig3_c}]{\includegraphics [width=0.990\linewidth]{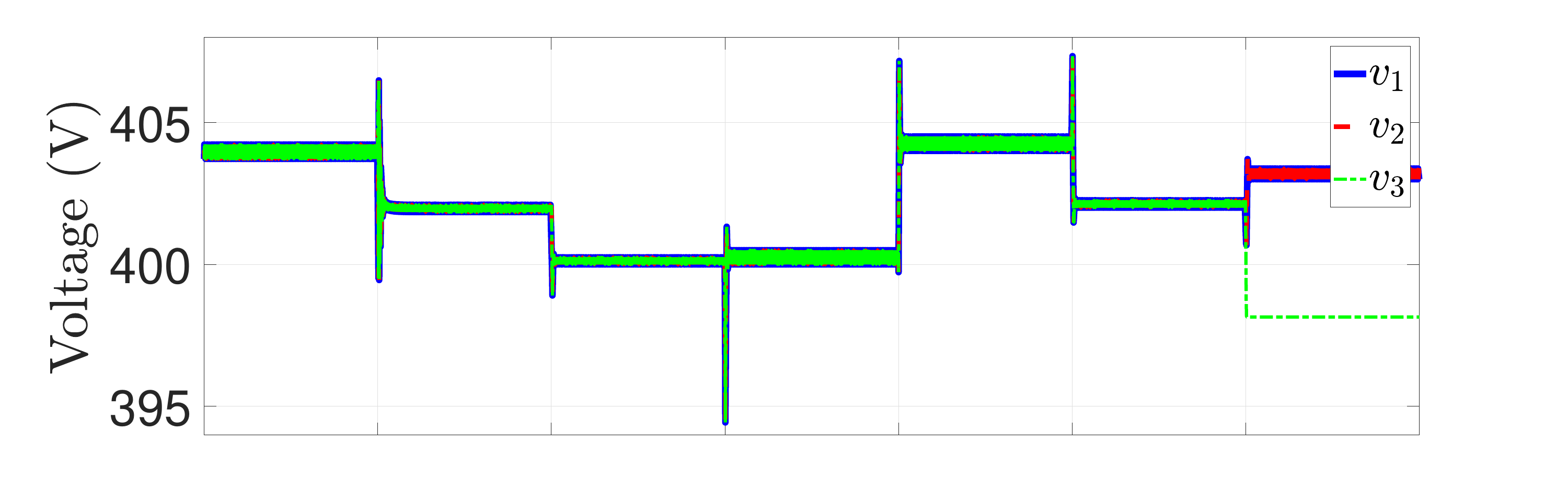}}\\
    \subfloat[\label{fig3_d}]{\includegraphics [width=0.990\linewidth]{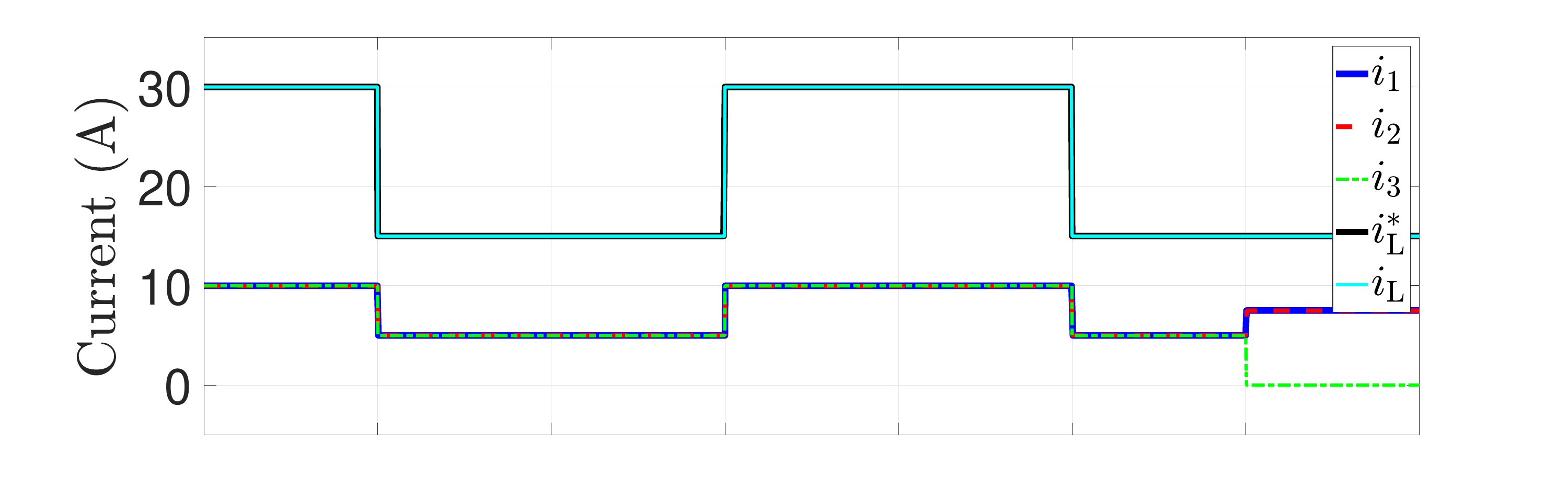}}\\ 
    \subfloat[\label{fig3_e}]{\includegraphics [width=0.990\linewidth]{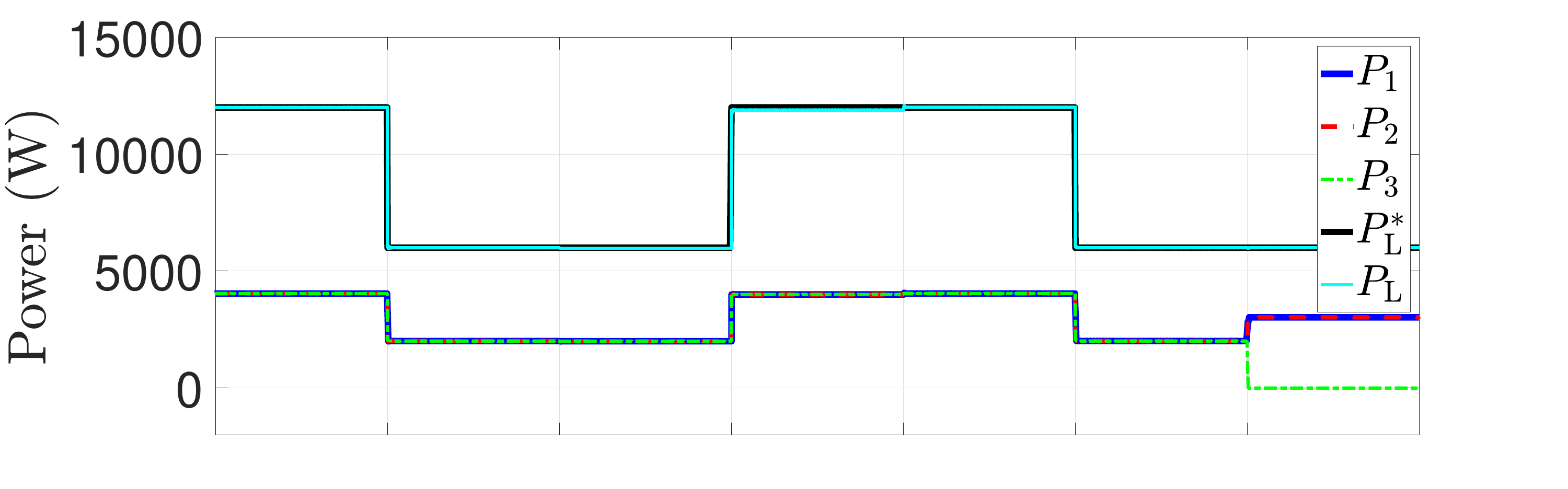}}\\
    \subfloat[\label{fig3_f}]{\includegraphics [width=0.990\linewidth]{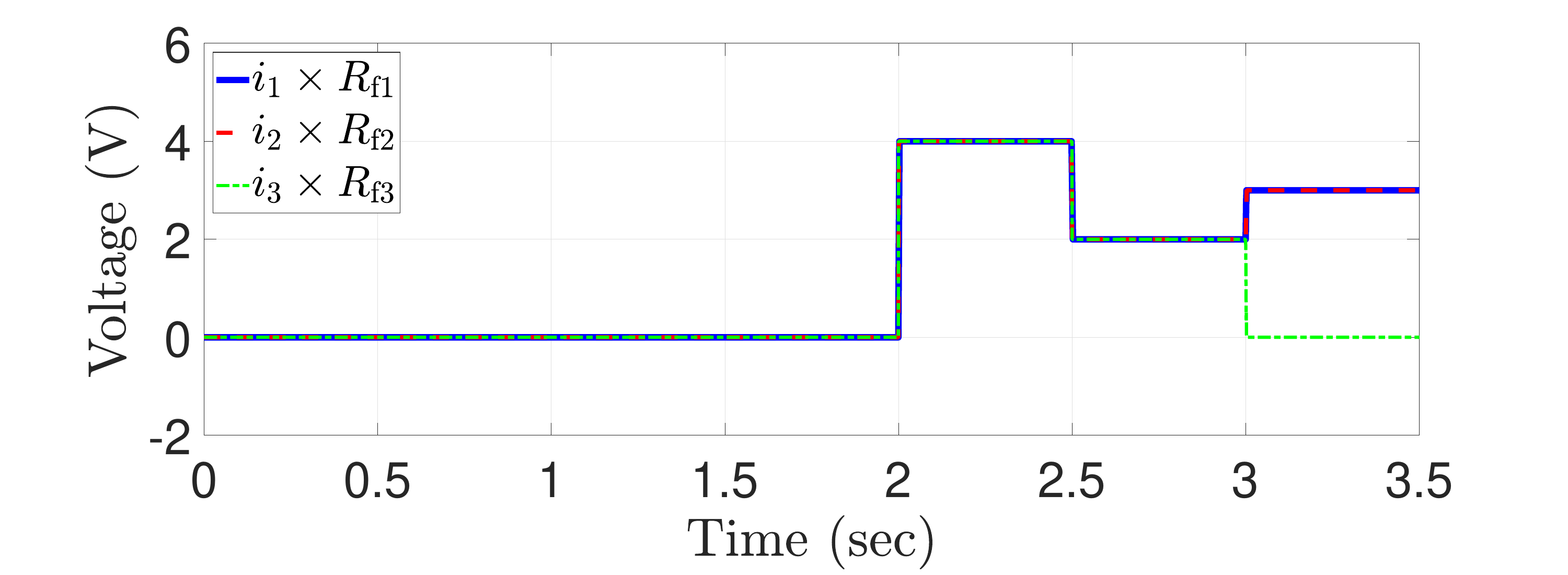}}\\
\caption{Simulation results of the DC microgrid with three boost converters under the different control mode strategies and load variations: 
(a) control mode, 
(b) microgrid voltage, 
(c) terminal voltages of the converters,
(d) terminal current of the converters,
(e) terminal power of the converters,
(f) voltages generated by the proposed decentralized control approach.} 
\label{fig_Sim_Res}
\vspace{-0.350cm}
\end{figure}
\section{Simulation Results}\label{Sec_4}
In this section, simulation tests are conducted to evaluate DC bus voltage restoration in DC microgrids using both traditional centralized and proposed decentralized control approaches. The DC microgrid, featuring three identical boost converters, is simulated in MATLAB/Simulink with the parameters listed in Table~\ref{table_1}.

Three case studies are considered: evaluating the traditional centralized control (with the secondary control), examining a base case without secondary control, and assessing the proposed decentralized control. Moreover, in each of the aforementioned case studies, the microgrid load is varied between full load (30 A) and half load (15 A) to observe the performance capability under these control strategies.

Fig.~\ref{fig_Sim_Res} presents the simulation results of the DC microgrid under the investigated control strategies. Fig.~\ref{fig_Sim_Res}(a) shows the control modes of the microgrid. From 0 s to 1 s, the traditional centralized control approach, depicted in Fig.~\ref{Fig_Meth_Trad_cont}, is enabled. Subsequently, from 1 s to 2 s, the traditional centralized control approach is disabled, followed by the activation of the proposed decentralized control approach, depicted in Fig.~\ref{Fig_Meth_Prop_cont}, from 2 s to 3.5 s.

Fig.~\ref{fig_Sim_Res}(b) presents the microgrid voltage, displaying both the desired reference value of $v^*_\mathrm{dc} = 400$ V and the measured value, $v_\mathrm{pcc}$. It is noted that under the traditional centralized control approach, the DC voltage is accurately restored to the desired value before and after the load step at 0.5 s. However, upon disabling the secondary control between 1 s and 2 s, $v_\mathrm{pcc}$ decreases from 400 V to around 398 V for a 15 A load current and further to around 396 V for a 30 A load current. Subsequently, the voltage is restored to the nominal value upon enabling the proposed decentralized control approach between 2 s and 3.5 s. The reliability of the proposed control is evident as it continues to regulate the DC bus voltage to the nominal value before and after the load current step between 2 s and 3 s, and after deactivating the third boost converter at 3 s.

Fig.~\ref{fig_Sim_Res}(c) displays the terminal DC voltage of the three boost converters. Overall, the terminal voltages of the boost converters are higher than the nominal voltage value. This occurs under both the traditional centralized control approach and the proposed decentralized control approach in order to compensate for voltage drops across the resistances of the feeders, ensuring the DC bus voltage is regulated to the nominal value of 400 V. Additionally, these voltages are nearly identical for all tested cases, except for the terminal voltage of converter 3, $v_3$, when disconnected for $t > 3$ s.

Fig.~\ref{fig_Sim_Res}(d) displays the output current of the converters along with the delivered total load measured and reference currents. It is noteworthy that the microgrid load type is DC current, ensuring the delivered current to the load remains constant regardless of DC bus voltage variations under the three tested control strategies. The total load current varies between 30 A and 15 A for each control strategy, with all boost converters delivering the same amount of current. Additionally, from 3 s to 3.5 s, the current is equally shared between converters 1 and 2 after converter 3 is switched off, causing its output current to drop to zero.

Fig.~\ref{fig_Sim_Res}(e) presents the corresponding output power of the three converters as well as the total power delivered to the load and its reference value. Considering the nominal voltage of 400~V and current reference values of 30 and 15~A, the load power reference varies between 12 and 6 kW. At each time instance, all three converters supply the same amount of 4 kW before load reduction, and approximately 2 kW afterward. However, between 3 and 3.5 s, when converter 3 is switched off, the supplied power by converters 1 and 2 increases from 2 kW to around 3 kW, accurately distributing the power previously supplied by converter 3.

Fig.~\ref{fig_Sim_Res}(f) displays the corresponding voltages generated by the proposed decentralized control approach. As depicted in Fig.~\ref{Fig_Meth_Prop_cont}, these voltages are generated locally to adjust the voltage reference for each converter and restore the DC bus voltage to the nominal value. Non-zero values of these voltages are observed after enabling the proposed control approach at 2 s. They are automatically calculated and adjusted with the output current of the converters, as demonstrated at 2.5 s when the load current changes from 30 A to 15 A. Additionally, the generated voltages of converters 1 and 2 increase at $t > 3$ s when converter 3 is disconnected.

\section{Hardware-in-the-Loop Experiments}\label{Sec_4}
\begin{figure*}[!t]
  \centering
  \subfloat[\label{fig4_a}]{\includegraphics [width=0.720\linewidth]{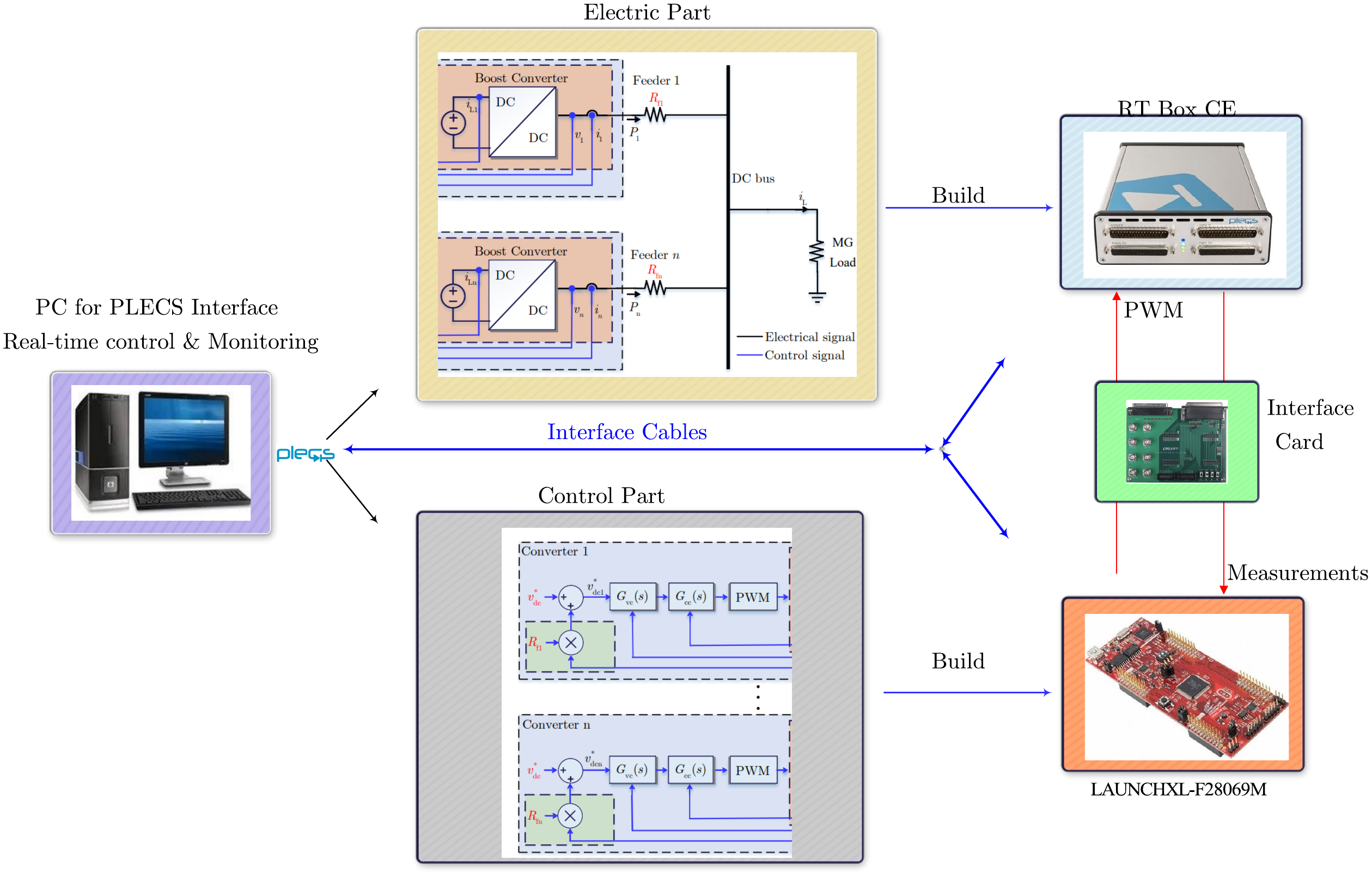}} \\ 
  \subfloat[\label{fig4_b}]{\includegraphics [width=0.550\linewidth]{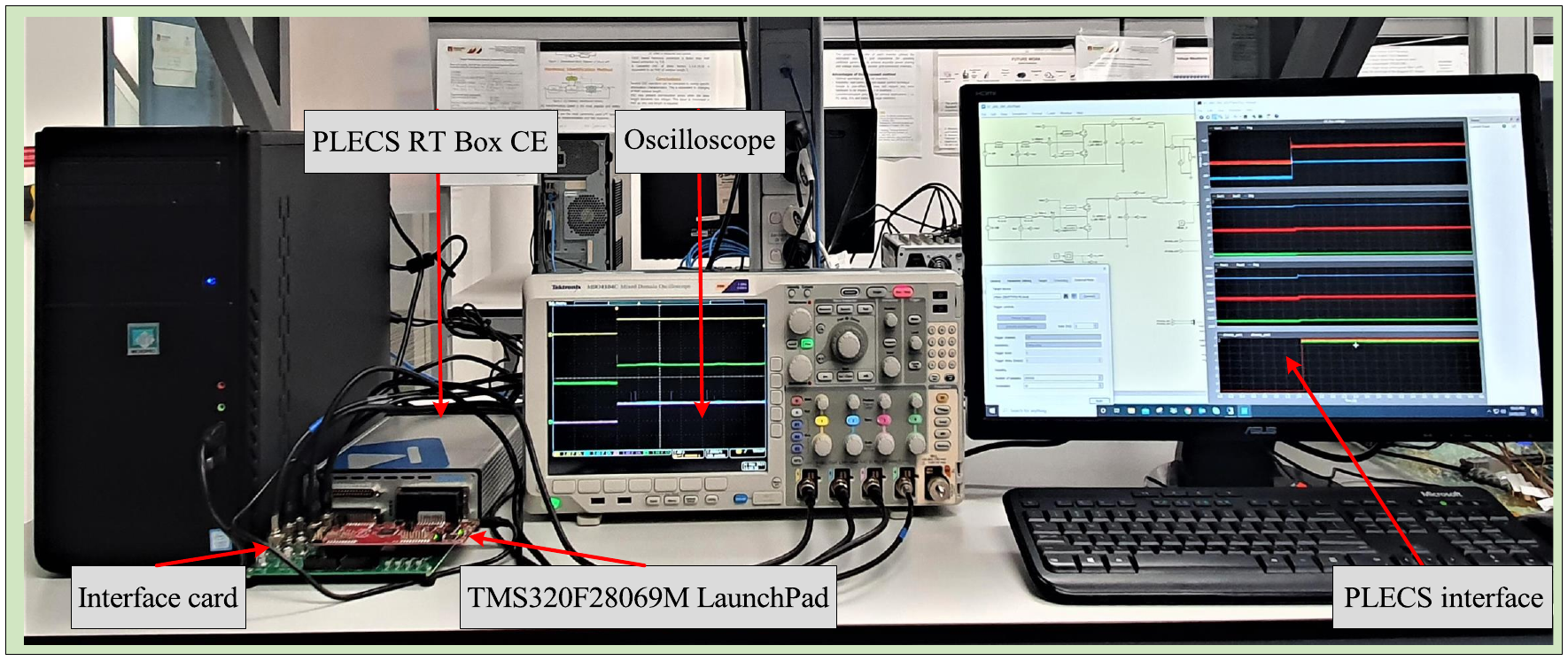}}\\
   \caption{Hardware-in-the-loop setup: (a) Block diagram of the setup, (b) Screenshot of the setup.} 
  \label{fig_ExpHiL_fig_0}
\vspace{-0.350cm}
\end{figure*}
\begin{figure}[!t]
  \centering
  \subfloat[\label{fig4_a}]{\includegraphics [width=0.8600\linewidth]{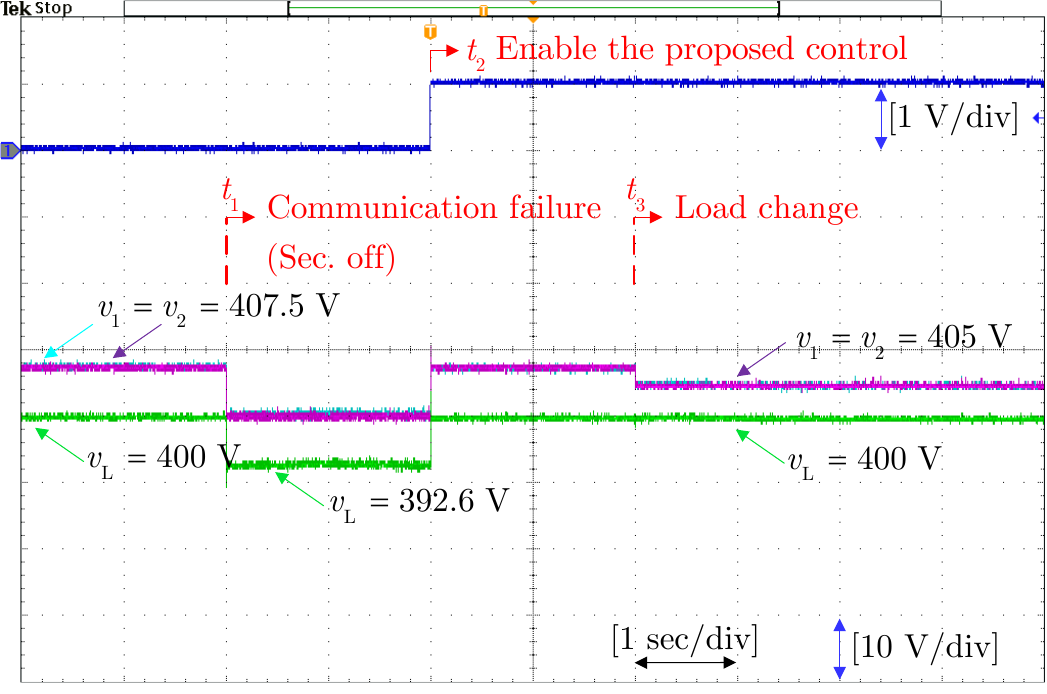}} \\ 
  \subfloat[\label{fig4_b}]{\includegraphics [width=0.8600\linewidth]{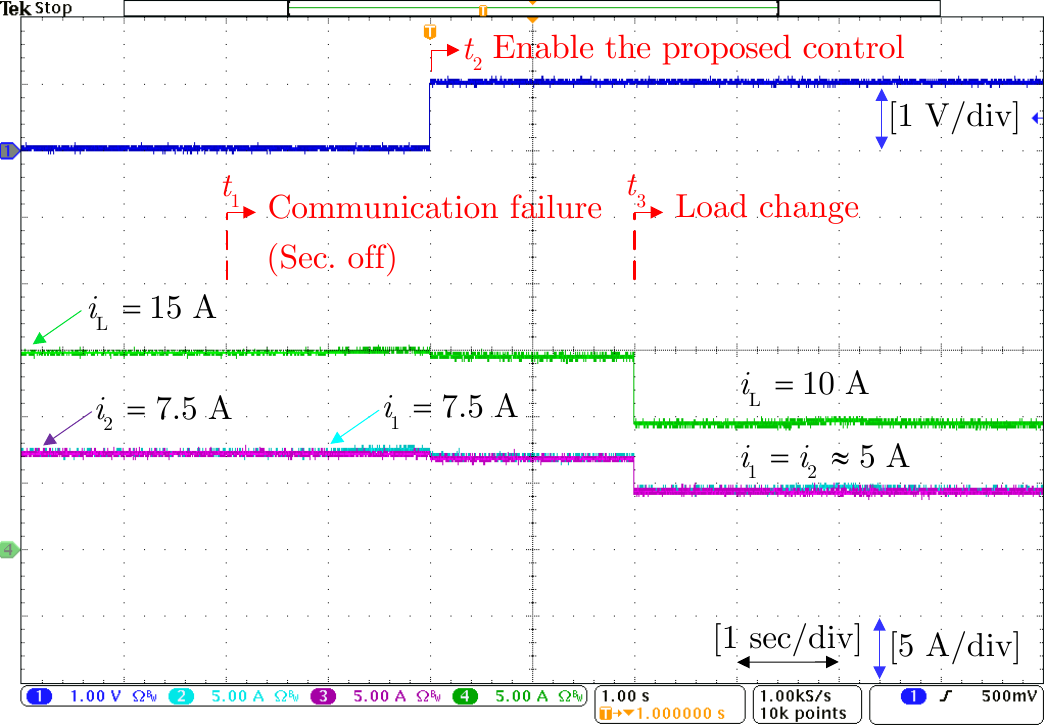}}\\
   \subfloat[\label{fig4_c}]{\includegraphics [width=0.8600\linewidth]{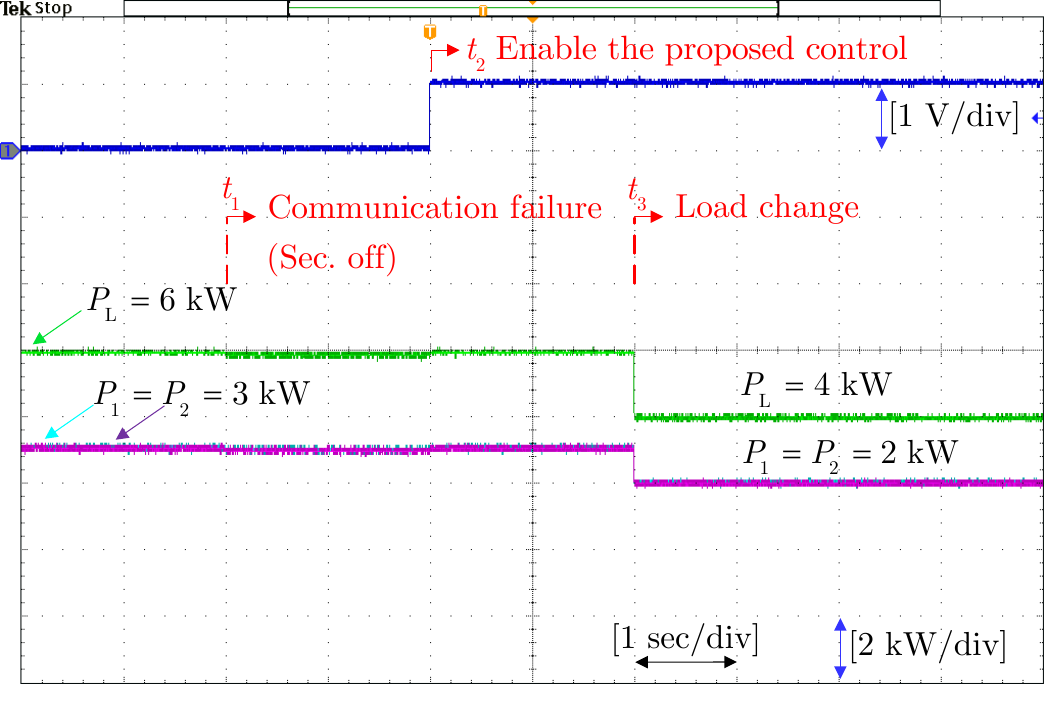}}\\
   \caption{
   HIL results of the DC microgrid with two boost converters under the different control mode strategies and load variations: (a) Voltages, (a) Currents, (b) Powers.} 
  \label{fig_Hil_Res}
\vspace{-0.350cm}
\end{figure}
To validate the proposed decentralized DC voltage restoration control approach, the system depicted in Fig.~\ref{Fig_Meth_Prop_cont} is considered for hardware-in-the-loop (HIL) experiments with two identical boost converters. The system parameters for the HIL validation are identical to those listed in Table.~\ref{table_1}, except that the initial microgrid load is 15 A, which also reduced to 10 A. The HIL setup, as shown in Fig.~\ref{fig_ExpHiL_fig_0}, involves deploying the power stage (electric part) of the microgrid into the PLECS RT Box from PLEXIM, while the control algorithm is implemented in real-time using the TMS320F28069M LaunchPad from Texas Instruments.

The performance of the microgrid is investigated in four scenarios. In the first scenario, prior to $t_1$, the secondary control is enabled for the DC bus voltage restoration of the microgrid. In the second scenario, at $t_1$, the secondary control is disabled (switched off), representing the loss of communication links between the common MGCC and the local boost converter controllers. The third scenario starts at $t_2$, where the proposed control is enabled for the DC bus voltage restoration of the microgrid. Finally, in the fourth scenario, starting at $t_3$, the proposed control is tested under load change.

Fig.~\ref{fig_Hil_Res}(a) displays the microgrid voltage at the PCC as well as the terminal output voltages of the two converters. Initially, the microgrid voltage was maintained equal to the reference value of 400 V when the secondary control was enabled. However, after the disabling of the secondary control at $t_2$, the microgrid voltage dropped to 392.6 V. Subsequently, the proposed control restored the microgrid voltage to its reference value when enabled at $t_2$, even after the load change at $t_3$. 
Fig.~\ref{fig_Hil_Res}(b) depicts the load current waveform along with the current delivered by the two converters to the PCC. Initially, the 15 A load current was equally shared by the two identical converters. Following the load change at $t_3$ to 10 A, each converter supplied 5 A to the load. Again, the maintenance of equal shared current between the two converters was observed before and after the enabling of the proposed control approach at $t_2$. 
Similarly,  the load power waveform and the power delivered by the two converters to the PCC are presented in Fig.~\ref{fig_Hil_Res}(c). Initially, the 6 kW load was equally shared by the two identical converters. Following the load change at $t_3$ to 4 kW, each converter supplied 2 kW to the load. The maintenance of equal shared power between the two converters was observed both before and after enabling of the proposed control approach at $t_2$. It can be seen that there is a slight decrease in the delivered power to the load observed between $t_1$ and $t_2$, attributed to the decrease in the microgrid voltage.

Overall, the HIL results validate the proposed simple yet effective decentralized DC bus voltage restoration for microgrid applications, enhancing microgrid reliability against communication link failures and mitigating  potential cybersecurity threats associated with communication links.

\section{Conclusions}\label{Sec_5}
This paper presents a straightforward approach to restoring the voltage of the common DC bus in DC microgrids. The method ensures DC bus voltage regulation in a decentralized manner, making it suitable for fully decentralized DC microgrids. In the presented approached, Voltage drops across the resistances of feeders are compensated for by using an additional control loop feedback within each converter, based on the converter output current and feeder resistance. The efficacy, robustness, and adaptability of the proposed decentralized control strategy are validated through simulation and hardware-in-the-loop results under various scenarios, including those with and without secondary control, with and without the proposed control strategy, and under both load and voltage reference changes. This approach eliminates the need for communication links, enhancing the reliability of the microgrid system without requiring a secondary controller, and thereby reducing potential cybersecurity threats. Future work can explore similar approaches for hybrid DC/AC microgrid systems with non-identical parameters of converters and feeders, mesh configurations, as well as for restoring end voltage for off-board EV chargers.

\section*{Acknowledgment}
The research reported in this publication was supported by funding from King Abdullah University of Science and Technology (KAUST) - Center of Excellence for Renewable Energy and Storage Technologies (CREST), under award number 5937.

\ifCLASSOPTIONcaptionsoff
 \newpage
\fi
\bibliography{IEEEabrv,References}

\begin{thebibliography}{10}
\providecommand{\url}[1]{#1}
\csname url@samestyle\endcsname
\providecommand{\newblock}{\relax}
\providecommand{\bibinfo}[2]{#2}
\providecommand{\BIBentrySTDinterwordspacing}{\spaceskip=0pt\relax}
\providecommand{\BIBentryALTinterwordstretchfactor}{4}
\providecommand{\BIBentryALTinterwordspacing}{\spaceskip=\fontdimen2\font plus
\BIBentryALTinterwordstretchfactor\fontdimen3\font minus \fontdimen4\font\relax}
\providecommand{\BIBforeignlanguage}[2]{{%
\expandafter\ifx\csname l@#1\endcsname\relax
\typeout{** WARNING: IEEEtran.bst: No hyphenation pattern has been}%
\typeout{** loaded for the language `#1'. Using the pattern for}%
\typeout{** the default language instead.}%
\else
\language=\csname l@#1\endcsname
\fi
#2}}
\providecommand{\BIBdecl}{\relax}
\BIBdecl

\bibitem{hatziargyriou2007microgrids}
N.~Hatziargyriou, H.~Asano, R.~Iravani, and C.~Marnay, ``Microgrids,'' \emph{IEEE power and energy magazine}, vol.~5, no.~4, pp. 78--94, 2007.

\bibitem{pogaku2007modeling}
N.~Pogaku, M.~Prodanovic, and T.~C. Green, ``Modeling, analysis and testing of autonomous operation of an inverter-based microgrid,'' \emph{IEEE Transactions on power electronics}, vol.~22, no.~2, pp. 613--625, 2007.

\bibitem{guerrero2010hierarchical}
J.~M. Guerrero, J.~C. Vasquez, J.~Matas, L.~G. De~Vicu{\~n}a, and M.~Castilla, ``Hierarchical control of droop-controlled ac and dc microgrids—a general approach toward standardization,'' \emph{IEEE Transactions on industrial electronics}, vol.~58, no.~1, pp. 158--172, 2010.

\bibitem{mohammed2023accurateVOC}
N.~Mohammed, M.~Ali, M.~Ciobotaru, and J.~Fletcher, ``Accurate control of virtual oscillator-controlled islanded ac microgrids,'' \emph{Electric Power Systems Research}, vol. 214, p. 108791, 2023.

\bibitem{10542752}
G.~Krishnan~S and C.~Konstantinou, ``Design and evaluation of a dc microgrid testbed for der integration and power management,'' in \emph{2024 12th Workshop on Modeling and Simulation of Cyber-Physical Energy Systems (MSCPES)}, 2024, pp. 1--6.

\bibitem{dragivcevic2015dc}
T.~Dragi{\v{c}}evi{\'c}, X.~Lu, J.~C. Vasquez, and J.~M. Guerrero, ``Dc microgrids—part i: A review of control strategies and stabilization techniques,'' \emph{IEEE Transactions on power electronics}, vol.~31, no.~7, pp. 4876--4891, 2015.

\bibitem{nasirian2014distributed}
V.~Nasirian, A.~Davoudi, F.~L. Lewis, and J.~M. Guerrero, ``Distributed adaptive droop control for dc distribution systems,'' \emph{IEEE Transactions on Energy Conversion}, vol.~29, no.~4, pp. 944--956, 2014.

\bibitem{mohammed2023accurate}
N.~Mohammed, L.~Callegaro, M.~Ciobotaru, and J.~M. Guerrero, ``Accurate power sharing for islanded dc microgrids considering mismatched feeder resistances,'' \emph{Applied Energy}, vol. 340, p. 121060, 2023.

\bibitem{lu2013improved}
X.~Lu, J.~M. Guerrero, K.~Sun, and J.~C. Vasquez, ``An improved droop control method for dc microgrids based on low bandwidth communication with dc bus voltage restoration and enhanced current sharing accuracy,'' \emph{IEEE transactions on power electronics}, vol.~29, no.~4, pp. 1800--1812, 2013.

\bibitem{montegiglio2023decentralized}
P.~Montegiglio, G.~Acciani, M.~Dicorato, G.~Forte, and F.~Marasciuolo, ``A decentralized power and bus voltage regulation approach for dc microgrids,'' \emph{IEEE Transactions on Industry Applications}, 2023.

\bibitem{wang2024adaptive}
X.~Wang, J.~Huang, Y.~Cao, T.~Yang, Q.~Xu, C.~Zhang, and X.~Zhang, ``Adaptive voltage-guaranteed control of dc/dc-buck-converter-interfaced dc microgrids with constant power loads,'' \emph{IEEE Transactions on Industrial Electronics}, 2024.

\bibitem{zhou2020cyber}
Q.~Zhou, M.~Shahidehpour, A.~Alabdulwahab, and A.~Abusorrah, ``A cyber-attack resilient distributed control strategy in islanded microgrids,'' \emph{IEEE Transactions on Smart Grid}, vol.~11, no.~5, pp. 3690--3701, 2020.

\end{thebibliography}

\bibliographystyle{IEEEtran} 

\end{document}